# Both the validity of the cultural tightness index and the association with creativity/order are spurious – a comment on Jackson et al. (2019)

Alexander Koplenig[1] and Sascha Wolfer[1]

1 Leibniz-Institute for the German Language (IDS), Mannheim, Germany.

Correspondence: koplenig@ids-mannheim.de

It was recently suggested in a study published in *Nature Human Behaviour* that the historical loosening of American culture was associated with a trade-off between higher creativity and lower order[1]. To this end, Jackson et al. generate a linguistic index of cultural tightness based on the Google Books Ngram corpus[2] and use this index to show that American norms loosened between 1800 and 2000[1]. While we remain agnostic toward a potential loosening of American culture and a statistical association with creativity/order, we show here that the methods used by Jackson et al. are neither suitable for testing the validity of the index nor for establishing possible relationships with creativity/order.

Firstly, to demonstrate that the linguistic cultural tightness index captures meaningful trends that can be attributed to changing norm strength, Jackson et al. show that there are statistically significant correlations between the index and five convergent measures (religiosity, laws passed by the US Congress, Supreme Court cases, execution rates and profanity in American television shows). Based on this analysis, Jackson et al. conclude that the "linguistic indicator is valid" (p. 246). However, there is a natural ordering in the observations in a times series and not accounting for this temporal aspect of time series very often results in incorrect statistical inference[3–6]. To see why this is problematic for the analyses presented by Jackson et al., note that the cultural tightness index exhibits a clear downward trend (Kendall tau correlation coefficient with year $\tau_Y = -0.97$) and there is strong within-series dependence (correlation between the current and its lagged value $\tau_L = 0.98$). For the analysis of temporal data, this has important ramifications[5]. In particular, the Pearson correlation coefficient between two series will be high, even though they are not related in any substantial sense, when the two series are trending or when there is within-series dependence, so called spurious or non-sense correlations[3–5,7]. To show that this also the case for the Kendall correlation coefficient, we generated the following six sets of time series (mathematical details, code to reproduce our findings and visualizations for each set of series can be found at https://osf.io/fq2z7/?view_only=e39479c2c1534b41a27957521a4114c5):

- (i) 10,000 simulated white noise series;
- (ii) 10,000 simulated series with stochastic time trends, i.e. random walks with (negative) drift;
- (iii) 10,000 random walks without drift;
- (iv) 10,000 simulated series with linear deterministic time trends;
- (v) 10,000 simulated series with non-linear deterministic time trends;
- (vi) 24,897 time series that we extracted from the Our World in Data online platform (*OWID*) on a wide variety of topics such as economic development, living conditions, technology adoption, agricultural production or environmental change with available information for hundreds of different geographical regions, countries, socioeconomic factors and topic-dependent categories (e.g. different types of natural disasters).

As the cultural tightness index, all series except the white noise series show both pronounced trends over time (median absolute correlation $|\tau_Y|_{med} = 0.03$ for (i); $|\tau_Y|_{med} = 0.85$ for (ii); $|\tau_Y|_{med} = 0.44$ for (iii); $|\tau_Y|_{med} = 0.69$ for (iv); $|\tau_Y|_{med} = 0.86$ for (v) and $|\tau_Y|_{med} = 0.73$ for (vi)) and strong within-series dependence (median correlation $\tau_{Lmed} = -0.00$ for (i); $\tau_{Lmed} = 0.92$ for (ii); $\tau_{Lmed} = 0.85$ for (iii); $\tau_{Lmed} = 0.61$ for (iv); $\tau_{Lmed} = 0.82$ for (v); $\tau_{Lmed} = 0.87$ for (vi)).

We then calculated correlations between the cultural tightness index and each of the 74,897 series and extracted two-sided *P*-values. Plot (a) of Fig. 1 summarizes the results: all but one set show very strong signs of spuriousness as a significant Kendall correlation (at $P < 0.05$) is observed in at least 88% of all cases. The exception is the set of white noise series where – as theoretically expected – significant results at the nominal 5% level occur in only 4.6% of all cases. Since the series belonging to (i) – (v) are inherently random and since the full table of all results (available at https://osf.io/fq2z7/?view_only=e39479c2c1534b41a27957521a4114c5) clearly shows that the overwhelming majority of correlations for (vi) are undeniably spurious (e.g. a median Kendall correlation of -0.96 between cultural tightness and the number of landline telephone subscriptions in 115 different countries and geographical regions; all significant at $P < 0.001$), this analysis proves that Kendall correlations are not suited to test a potential relationship with tightness-looseness when the other involved series is trending or when there is within-series dependence. The validity analysis of Jackson et al. is spurious because, as pointed out by Jackson et al. themselves (p. 246), all five convergent measures except execution rates show clear trends across time. A visual inspection of Supplementary figure 1 of Jackson et al. also suggests that all five series show strong within-series dependence, correspondingly all $\tau_L$-values $> 0.68$.

For each measure, we used a model selection algorithm[8] to find the best ARIMA model for the errors of a dynamic regression model[9] where cultural tightness is included as a covariate. Only in one of five cases is the coefficient of cultural tightness significant (at $P < 0.05$).

Secondly, to account for the influence of time in subsequent analyses, Jackson et al. run ordinary least square regressions (*OLS*) where both the outcome and the cultural tightness index are predicted by time and several other predictors in order to use the residuals obtained from the models as input for subsequent Kendall correlation analyses. This implies that

Jackson et al. assume that the time series can be modelled as linear deterministic trends[9]. Correspondingly, plot (b) of figure 1 shows that the problem of spuriousness with the *OLS* approach is alleviated for the simulated series with linear deterministic time trends where significant results only occur roughly twice as often as expected at the 5% significance level. In all other cases but the white noise series, the problem persists. A visual inspection of the Jackson et al. data clearly demonstrates that a linear parameterization is inadequate for both the cultural tightness index and all seven measures of creativity/order because it induces new spurious time-series patterns in the data (see Supplementary figure 1)[10]. In addition, the resulting residuals show pronounced signs of within-series dependence (all $\tau_L$-values $> 0.40$).

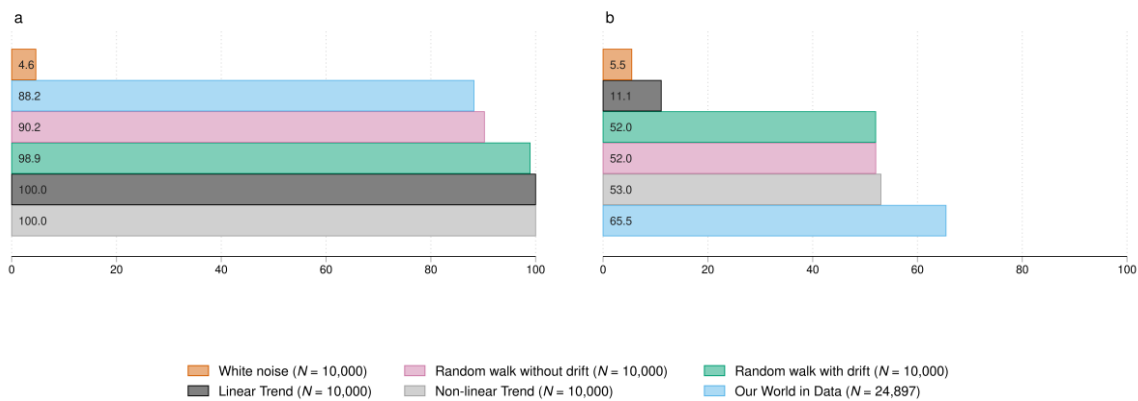

**Fig. 1 | Percentage of $P < 0.05$ for $\tau$ between cultural tightness and different types of time series. a,** without accounting for time. **b,** *OLS*-approach of Jackson et al. where the influence of time, collectivism and wealth is regressed out.

To highlight why such inappropriate transformations are highly consequential, we replicated the analyses of Jackson et al. in Table 1, column 2. While Jackson et al. interpret these results as evidence for a creativity-order trade off that can be linked to cultural tightness, we show in the subsequent columns of Table 1 that statistically significant patterns with similar or higher strengths are obtained if we replace the cultural tightness index by (a) the life expectancy in Costa Rica, (b) the worldwide aquaculture seafood production, (c) the population size of Uganda, (d) the number of tractors used in OECD member states, (e) the global palm oil production, or (f) the number of animal slaughtered for meat in India. It is worth noting that we are not just cherry-picking: of 24,549 *OWID* series with enough available information to test for a potential association with the 8 creativity/order variables, there are significant associations (at $P < 0.05$) with 6 out of 8 variables in almost 50% of all cases; all 8 associations are significant in 21.88% of all cases. This suggests that the evidence for an association between cultural tightness with a potential creativity-order trade off reported by Jackson et al. is solely the consequence of mis-specified models that do not correctly account for the underlying temporal structure of the data. Again, we used the model selection algorithm mentioned above to fit dynamic regression models where the cultural tightness index is included as a covariate. In none of the eight cases is cultural tightness a significant predictor of creativity/order (all $P$-values $> 0.10$).

**Table 1 | Replication of the association with creativity/order.** Column 2: replication of original results. Column 3 – 8: replication for several other predictors. * $P < 0.01$; ** $P < 0.005$

| Measure | Cultural tightness | Life expectancy Costa Rica | Seafood production worldwide | Population Uganda | Tractors OECD | Palm oil production worldwide | Slaughtered livestock India |
|---|---|---|---|---|---|---|---|
| Patent rates | -0.68** | -0.77** | 0.67** | 0.70** | -0.70** | 0.68** | -0.76** |
| Trademark rates | -0.62** | -0.71** | 0.83** | 0.78** | -0.63** | 0.81** | -0.67** |
| Feature film production | -0.41** | -0.46** | 0.44** | 0.50** | -0.38** | 0.41** | -0.38** |
| Baby-naming conformity | 0.56** | 0.53** | -0.49** | -0.65** | 0.46** | -0.44** | 0.55** |
| Household debt rates | -0.65** | -0.78** | 0.83** | 0.89** | -0.68** | 0.89** | -0.71** |
| Adolescent pregnancy rates | -0.43** | -0.42** | 0.31* | 0.37** | -0.41** | 0.36** | -0.41** |
| Crimes | 0.44** | 0.53** | -0.64** | -0.65** | 0.48** | -0.65** | 0.43** |
| High school enrolment | 0.49** | 0.67** | -0.67** | -0.70** | 0.54** | -0.65** | 0.58** |

**Data availability**

The original Jackson et al. dataset is available at https://osf.io/x2uzn/. All other time series were taken from https://ourworldindata.org and are described and cited at https://osf.io/fq2z7/?view_only=e39479c2c1534b41a27957521a4114c5.

**Code availability**

The code necessary to conduct the reported analysis is available in two different programming languages (Stata 14.2[11] and R[12]) on the Open Science Framework at https://osf.io/fq2z7/?view_only=e39479c2c1534b41a27957521a4114c5.

**Author contributions**

AK and SW designed and conceptualized the study and analysed the data. AK wrote the paper.

**Acknowledgments**

We thank Joshua Conrad Jackson for responding to our early notification of our intent to submit this report. We thank Michael E.W. Varnum and Igor Grossmann for suggesting the model selection algorithm to us and Sarah Signer for proofreading.

**Competing interests**

The authors declare no competing interests.

# Supplementary figure 1

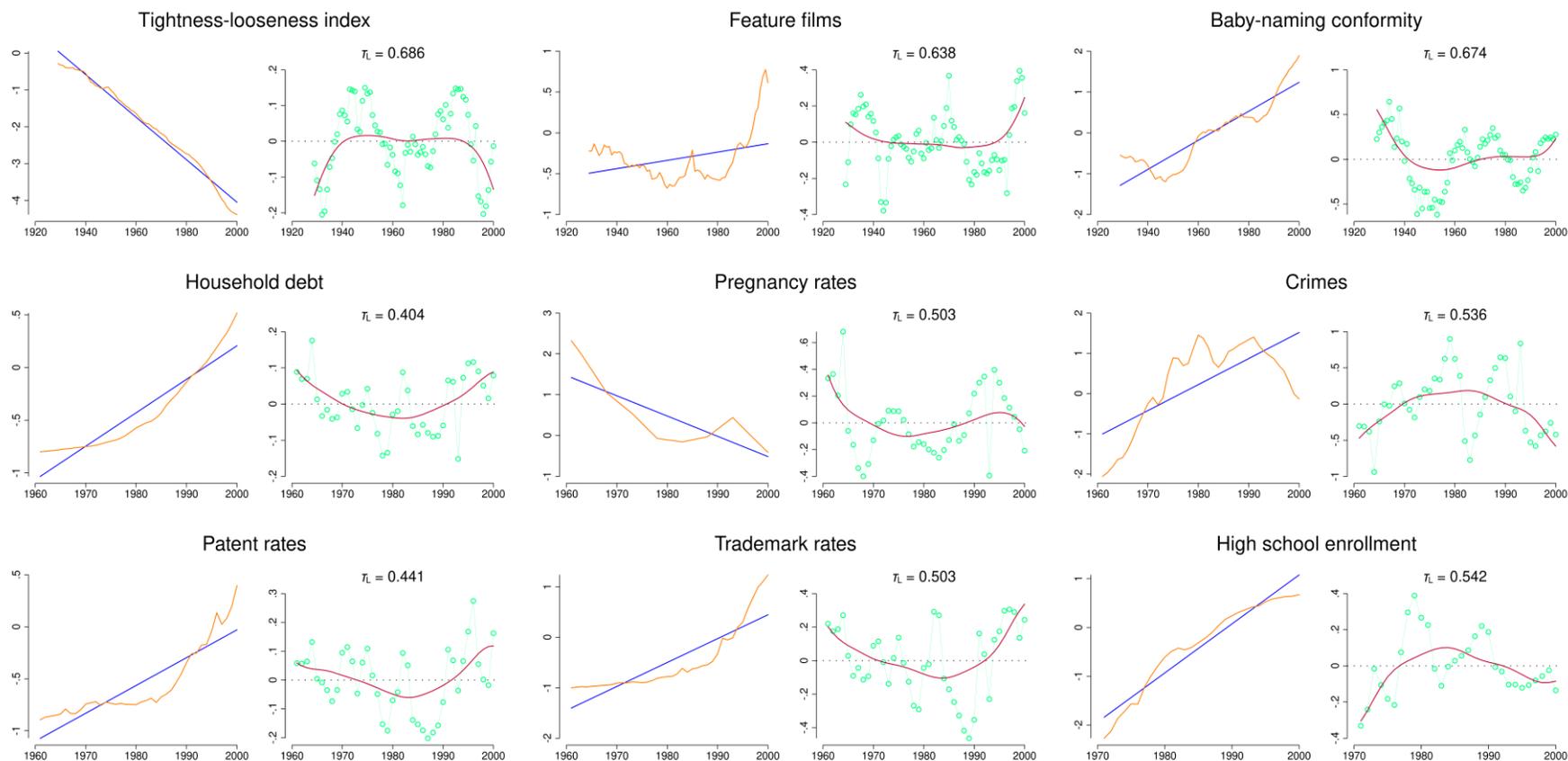

**Supplementary figure 1 | Visual analysis of linearity and within-series dependence for the cultural-tightness index and all eight measures of creativity/order.** Orange line in left-hand plots: observed levels; blue line: linear prediction. Connected mint circles in right-hand plots: residuals from the *OLS*-approach of Jackson et al[1]; cranberry line: loess smoothed lines. Note that the residuals take "extended excursions above and below 0"[2] which is indicative of within-series dependence; the $\tau_L$ values represent the Kendall correlation between current and lagged values for each residual. Further note the clear similarity

between the orange and the cranberry curves for each variable demonstrating that the linear detrending strategy of Jackson et al[1] induces new spurious time-series patterns in the data[3].